\documentclass[pra, twocolumn, showpacs, amsmath, amssymb , noshowkeys]{revtex4}

\usepackage{dcolumn}
\usepackage{bm}
\RequirePackage[usenames,dvipsnames]{color}

\usepackage{color}
\usepackage{epsfig} 
\psfull                                
\usepackage{mathbbol}
\usepackage{bbm}
\usepackage{dsfont}

 \usepackage{hyperref}

\newcommand{\bra}[1]{\langle #1|}	
\newcommand{\ket}[1]{|#1\rangle}

\newcommand{\rme}{\text{e}}
\newcommand{\rmi}{\text{i}}
\newcommand{\sign}[1]{\text{sign}\left(#1 \right)}

\newcommand{\eps}{\varepsilon}
\newcommand{\hh}{\frac{\hbar}{2}}

\newcommand{\Mixangle}{\Theta_{j}^l}
\newcommand{\sinmixa}{\sin \frac{\Mixangle}{2}}
\newcommand{\cosmixa}{\cos \frac{\Mixangle}{2}}
\newcommand{\half}{\frac{1}{2}}
\newcommand{\DXi}[2]{\Xi_{#2}^{#1}}

\newcommand{\Min}[1]{\text{Min} \left \{ #1 \right\} }
\newcommand{\ketU}[1]{\ket{\widetilde{\uparrow,#1}}}
\newcommand{\braU}[1]{\bra{\widetilde{\uparrow,#1}}}
\newcommand{\ketD}[1]{\ket{\widetilde{\downarrow,#1}}}
\newcommand{\braD}[1]{\bra{\widetilde{\downarrow,#1}}}

\begin{document}

    \title{Qubit-oscillator system: An analytical treatment of the ultrastrong coupling regime}
\date{\today}
\author{Johannes Hausinger} 
\email{johannes.hausinger@physik.uni-r.de}
\author{Milena Grifoni}
\affiliation{Institut f\"{u}r Theoretische Physik, Universit\"at
Regensburg, 93040 Regensburg, Germany}

\begin{abstract}
  We examine a two-level system coupled to a quantum oscillator, typically representing experiments in cavity and circuit quantum electrodynamics. We show how such a system can be treated analytically in the ultrastrong coupling limit, where the ratio $g/\Omega$ between coupling strength and oscillator frequency approaches unity and goes beyond. In this regime the Jaynes-Cummings model is known to fail, because counter-rotating terms have to be taken into account. 
By using Van Vleck perturbation theory to higher orders in  the qubit tunneling matrix element $\Delta$ we are able to enlarge the regime of applicability of existing analytical treatments, including in particular also the finite bias case.
We present a detailed discussion on the energy spectrum of the system and on the dynamics of the qubit for an oscillator at low temperature. We consider the coupling strength $g$ to all orders, and the validity of our approach is even enhanced in the ultrastrong coupling regime. Looking at the Fourier spectrum of the population difference, we find that many frequencies are contributing to the dynamics. They are gathered into groups whose spacing depends on the qubit-oscillator detuning. Furthermore, the dynamics is not governed anymore by a vacuum Rabi splitting which scales linearly with $g$, but by a non-trivial dressing of the tunneling matrix element, which can be used to suppress specific frequencies through a variation of the coupling.
\end{abstract}
\pacs{03.67.Lx, 42.50.Pq, 85.25.Cp}
\keywords{}
\maketitle

\section{Introduction}
The model of a two-level system coupled to a quantized oscillator experiences widespread application in many different fields of physics. In quantum optics it describes the interaction of light with matter -- of an atom  coupled to the electromagnetic mode of a cavity. Most interesting in this instance is the regime of strong coupling; i.e.,  the coupling strength $g$ between the atom and the cavity mode exceeds the loss rates stemming from spurious processes like escape through the cavity mirrors, relaxation to other atomic levels or into different photon modes, or decay due to fluctuations in the qubit control parameter induced by the environment. Under this condition, the atom and the cavity can repeatedly exchange excitations before decoherence takes over. The resulting Rabi oscillations have been observed experimentally and the field is known today as cavity quantum electrodynamics \cite{Raimond2001, Mabuchi2002}. But also for artificial atoms, like superconducting qubits \cite{Makhlin2001, You2005, Clarke2008}, similar setups have been realized with the cavity being formed by a one-dimensional transmission line resonator \cite{Blais2004, Wallraff2004} or a simple $LC$-circuit \cite{Chiorescu2004, Johansson2006}. In both cases the Rabi splitting in the qubit-oscillator spectrum could be detected \cite{Wallraff2004, Chiorescu2004}, while in the experiment of Johansson et al. \cite{Johansson2006} coherent vacuum Rabi oscillations were observed. The advantages of this field, known as  circuit quantum electrodynamics (QED), are manifold: For instance, the transition dipole moment of a superconducting Cooper-pair box can be made up to four orders of magnitude larger than in real atoms. Using a coplanar waveguide as cavity, the volume can be confined very tightly in the transverse directions only limited by the qubit size, which can be made much smaller than the resonator wavelength. Thus, we can speak of a quasi-1D cavity, which leads to a strongly enhanced electric field \cite{Blais2004, Wallraff2004} and the strong coupling limit is more easily reached. In the first realization of Wallraff et al. \cite{Wallraff2004} a coupling strength of $g/\Omega \sim 10^{-3}$ was observed, while in more recent experiments couplings up to a few percent, $g/\Omega \lesssim 0.025$, were reported \cite{Sillanpaa2007, Majer2007, Fink2008, Deppe2008, Bishop2009}, reaching the upper limit possible for electric dipole coupling \cite{Devoret2007, Schoelkopf2008}, whereas in cavity QED one finds typically $g/\Omega \sim 10^{-6}$ \cite{Raimond2001}. The artificial atom can be placed at a fixed location in the cavity, so that fluctuations in the coupling strength are avoided. Furthermore, fabrication techniques known from integrated circuits can be used to ``wire-up'' the qubit cavity system and connect it to other circuit elements \cite{Schoelkopf2008}.
For investigations on the qubit-oscillator setup, the Jaynes-Cummings model (JCM) \cite{Jaynes1963} is usually invoked. It relies on a rotating-wave approximation (RWA), which is valid for not too strong coupling $g \ll \Delta_\text{b}, \Omega$ and weak detuning, $\Delta_\text{b} \approx \Omega$, where  the qubit transition frequency $\Delta_\text{b} = \sqrt{\eps^2 + \Delta^2}$ equals the tunneling matrix element $\Delta$ for zero static bias $\eps$.
However, for certain experimental conditions, coupling strengths of more than a few percent or even unity were predicted reaching the ultrastrong coupling regime \cite{Ciuti2005, Devoret2007, Schoelkopf2008, Bourassa2009}. For those strong couplings, the application of a rotating-wave approximation  and thus the JCM is \textit{not} justified anymore. For instance, quite recently an experiment by Niemczyk et al. \cite{Niemczyk2010} could show the failure of the JCM for a Josephson flux-qubit placed inside the center conductor of an inhomogeneous transmission-line resonator. Also for a flux-qubit coupled to an $LC$-circuit, the break-down of the rotating wave approximation has been demonstrated experimentally \cite{Forn-Diaz2010} and the ultrastrong coupling regime seems to be in close reach \cite{Fedorov2010}.  While in the JCM the ground state of the qubit-oscillator system consists of a product of the qubit's ground state and the oscillator's vacuum state, an inclusion of the counter-rotating terms leads to -- depending on the coupling strength -- an entangled or a squeezed vacuum state containing virtual photons \cite{Ciuti2005, Ashhab2010}, which under abrupt switch-off of the coupling are emitted as correlated photon pairs, reminding of the dynamical Casimir effect \cite{Ciuti2005, Guenter2009, Anappara2009}. Such an adiabatic ma\-ni\-pu\-la\-tion has been recently realized experimentally for intersubband cavity polaritons in semiconducting quantum wells \cite{Guenter2009}. In this experiment and also in \cite{Anappara2009} a dimensionless coupling strength of about 10\% has been reached. Furthermore, ultrastrong coupling has been predicted for qubits coupled to nanomechanical resonators \cite{Irish2005}.\newline
Theories examining the qubit-oscillator system going beyond the RWA are at hand: The adiabatic approximation (see \cite{Irish2005} and references therein) relies on a polaron transformation and is derived under the assumption $\Omega \gg \Delta_\text{b}$. It fails to return the limit of zero coupling $g \to 0$, where the JCM works well. An improvement to this theory is given by the generalized rotating-wave approximation (GRWA) \cite{Irish2007}, which is a combination of the adiabatic approximation and the standard RWA  and works  well in both regimes of zero and large qubit-oscillator detuning. Further, it covers correctly the weak coupling limit. However, it has not been used yet to investigate the dynamics of the qubit-oscillator system. The NIBA calculations by Nesi et al. \cite{Nesi2007} treat analytically a two-level system coupled to a harmonic oscillator to all orders in the coupling strength $g$, taking environmental influences into account. Zueco et al. present a theory beyond the rotating-wave approximation in the strong dispersive regime \cite{Zueco2009}. From these works, one can learn that the simple picture of the qubit-oscillator energy spectrum is not given by the Jaynes-Cummings ladder anymore, where pairs of energy levels which are degenerate for $g=0$ are split by $2g \sqrt{j}$, with $j$ denoting the higher oscillator level being involved. However, all these theories are derived for an unbiased qubit ($\eps=0$) or in the terminology of cavity and circuit QED, for a qubit operated at the degeneracy point or sweet spot. While this situation is usually encountered for real atoms in cavity QED,  it is quite straightforward to vary the static bias $\eps$ of superconducting qubits by an external control parameter like the gate voltage applied to a Cooper-pair box or the magnetic flux acting on a Josephson junction. Indeed, such a detuning from the degeneracy point is performed in spectroscopic measurements of the qubit-oscillator system, see e.g. \cite{Wallraff2004, Forn-Diaz2010}, or in a current-based read-out of the qubit \cite{Chiorescu2003}.
Therefore, theories are necessary which treat the \textit{biased} qubit-oscillator system in the ultrastrong coupling limit. In \cite{Hausinger2008, Vierheilig2009} this is done for a qubit coupled to a linear or nonlinear oscillator, respectively, up to second order in the coupling strength $g$. Higher-order effects like the Bloch-Siegert shift of the qubit dynamics could be observed. Brito et al. used in \cite{Brito2008(2)} a slightly changed polaron transformation on the qubit-oscillator model and obtained by truncating the displaced harmonic oscillator to its first excited level an effective four-level model. Quite recently, the adiabatic approximation for a high-frequency oscillator was reviewed for a biased system \cite{Ashhab2010}. Furthermore, the opposite regime of a high-frequency qubit has been examined there. \newline
In this work, we present a theory which takes the static bias of the qubit into account and treats the qubit-oscillator system to all orders in the coupling strength. We consider the qubit tunneling matrix element $\Delta$ as a small perturbation. For zero static bias, our approach can be seen as an extension of the adiabatic approximation by taking into account higher order terms of $\Delta$ using Van Vleck perturbation theory (VVP). We do not only examine the energy levels of the system but also calculate corrections to the displaced qubit-oscillator states, which we obtain using a polaron transformation on the unperturbed ($\Delta = 0$) case. Unlike in the adiabatic approximation discussed in \cite{Ashhab2010}, we take the qubit's static bias into account while identifying degenerate subspaces, thereby adjusting the renormalized frequency already in the first order approach. Our results work very well for negative detuning ($\Delta_\text{b} < \Omega$) for the whole range of coupling strength and even exceeds in accuracy results obtained from the GRWA for $\eps =0$. For not too weak coupling $g/\Omega \gtrsim 0.5$ and/or finite static bias, it agrees with numerical results even for the resonant case $\Delta_\text{b} = \Omega$ or positive detuning  $\Delta_\text{b}> \Omega$. With these observations we believe we can close the gaps which cannot be treated  by the Jaynes-Cummings model or the GRWA. \newline
 With our investigations we enter a new physical regime: the splitting between the energy levels does not scale linearly in $g$ anymore but depends through a dressing by Laguerre polynomials on the coupling strength. This dependence  allows for a suppression of individual frequency contributions to the dynamics. We further discover that even at low temperatures several frequencies come into play, while the JC dynamics is usually governed by two main oscillations.\newline
The outline of this work is as follows: After introducing the Hamiltonian of the qubit-oscillator system in Sec. \ref{Sec::Hamiltonian}, we explain how it can be approximately diagonalized by a combination of displaced oscillator states and VVP. The resulting eigenstates and eigenenergies are given in Sec.  \ref{Sec::Diagonalization} being valid for the zero and nonzero bias case. For both situations, we examine the energy spectrum in detail in Sec. \ref{Sec::EnergySpec}, comparing the different approaches to numerical calculations. In Sec. \ref{Sec::Dynamics}, we concentrate on the dynamics; i.e., we determine the time evolution of the population difference of the two-level system and test the adiabatic approximation and VVP again against numerics.  We conclude our discussion in Sec. \ref{Sec::Conclusion}.

\section{Diagonalization of the qubit-oscillator Hamiltonian}
\subsection{The two-level-oscillator Hamiltonian} \label{Sec::Hamiltonian}
The predominant model to describe the interaction between an atom and the field of a cavity is the two-level-oscillator Hamiltonian, see, e.g., \cite{CohenTannoudji2004},
\begin{equation} \label{HTot}
 H= H_\text{TLS} + H_\text{int} + H_\text{osc}.
\end{equation}  
The atom is described as a simple two-level system (TLS),
\begin{equation} \label{HTLS}
 H_\text{TLS} = -\frac{\hbar}{2} (\eps \sigma_z + \Delta \sigma_x),
\end{equation}
where we use as basis the so-called localized states, which are eigenstates of the $\sigma_z$ Pauli matrix, $\sigma_z \ket{\uparrow / \downarrow} = \pm \ket{\uparrow / \downarrow}$. Tunneling between the two states is taken into account by $\Delta \sigma_x$ \footnote{We assume $\Delta \geq 0$ throughout this work.}, and $\eps$ describes a possible static bias of the TLS. In cavity QED setups one typically finds the situation of zero static bias, while in circuit QED $\eps$ can be controlled in situ. The atom is connected to the field of the cavity via a dipole coupling, which is expressed by  
\begin{equation}  \label{HInt}
 H_\text{int} = \hbar g \sigma_z (b^\dagger + b).
\end{equation} 
The coupling strength is given by $g$, while $b^\dagger$ and $b$ are the raising and lowering operators of the field. As usual, we assume that this field can be expressed by a single harmonic oscillator mode of frequency $\Omega$,
\begin{equation} \label{HOsc}
 H_\text{osc} = \hbar \Omega b^\dagger b,
\end{equation} 
where we neglected the zero-point energy. Despite its simplicity, this Hamiltonian cannot be diagonalized analytically, and several approximation schemes have been developed. The most famous one is the Jaynes-Cummings model \cite{Jaynes1963}, which neglects ``energy non-conserving'' or counter-rotating terms, and is restricted to relatively weak coupling strengths $g \ll \Delta_\text{b}, \Omega$, where $\Delta_\text{b} = \sqrt{\eps^2 + \Delta^2}$, and to systems close to resonance, $\Delta_\text{b} \approx \Omega$. A natural extension to the Jaynes-Cummings model (JCM) is given in \cite{Hausinger2008}, where the counter-rotating terms in the Hamiltonian (\ref{HTot}) are taken into account by using VVP to second order in the qubit-oscillator coupling. This method thus works also for intermediate coupling strengths and  biased qubits and is able to explain effects which go beyond the capabilities of the JCM like the Bloch-Siegert shift recently measured in \cite{Forn-Diaz2010}.  An approach which goes beyond the restriction of weak coupling is the ``adiabatic approximation in the displaced oscillator basis'', see \cite{Irish2005} and references therein. It is derived for the limit $\Omega \gg \Delta_\text{b}$ and relies on a separation of timescales: In order to calculate the fast dynamics of the oscillator (fast compared to the qubit), the part  coming from the TLS in Eq. (\ref{HTot}) is neglected, so that one gets an effective Hamiltonian for the oscillator reading
\begin{equation}
 \hbar g \sigma_z (b^\dagger + b) + \hbar \Omega b^\dagger b.
\end{equation}  
Thus, depending on the state of the qubit the oscillator is displaced in opposite directions, while not changing its energy for a fixed oscillator quantum $j$, as its eigenenergies are given by $\hbar j \Omega - \hbar g^2/\Omega^2$ \cite{Irish2005}. By reintroducing the qubit contribution this degeneracy is lifted. However, as long as $\Delta_\text{b} \ll \Omega$, the doublet structure is conserved. For an unbiased system, as done in \cite{Irish2005}, the condition translates to $\Delta \ll \Omega$ and the tunneling matrix element $\Delta$ can be treated as a small perturbation, in the end leading to an effective Hamiltonian consisting of 2 by 2 blocks, with a renormalized frequency on the off-diagonal. As this special case is included in our calculation, we will describe it in more detail below. Furthermore,  the contrary regime of a high-frequency qubit $\Delta_\text{b} \gg \Omega$ has been treated in \cite{Ashhab2010} analytically for certain special cases. This situation is also partly contained in our formalism.

\subsection{Eigenenergies and eigenstates} \label{Sec::Diagonalization}
In the following, we demonstrate how  the full Hamiltonian $H$ can be diagonalized perturbatively to second order in $\Delta$.
For a vanishing tunneling element, $\Delta = 0$, the polaron-like transformation 
\begin{equation}
  U = \rme^{g (b-b^\dagger) \sigma_z /\Omega}
\end{equation} 
brings $H$ into a diagonal form \footnote{In \cite{Brito2008(2)} it is pointed out that the simple polaron transformation fails in the limit of large tunneling elements $\Delta \gg \Omega$. For a flux-qubit this situation occurs for an applied external flux at which the qubit potential changes from a double-well to a single well, and thus the qubit eigenstates become delocalized. In our work, however, we do not aim at describing such a parameter regime.}.
Its eigenstates are $\ket{\widetilde{\uparrow /\downarrow, j}} = U \ket{\uparrow /\downarrow, j}$, where $\ket{\uparrow /\downarrow, j}$ are the eigenstates of the qubit-oscillator system for $\Delta=0$ and $g=0$. For detailed expressions see Eqs. (\ref{App::Delta0stateUp}) and (\ref{App::Delta0stateDown}). They correspond to the displaced oscillator states used in \cite{Irish2005}, where the displacement depends on the qubit state. The eigenvalues are 
\begin{equation}
 E_{\uparrow / \downarrow, j}^{0} = \mp \hh \eps + \hbar j \Omega -\hbar \frac{g^2}{\Omega}.
\end{equation} 
For finite $\Delta$, the perturbative matrix elements become \cite{Irish2005, Eckardt2008, Ashhab2010}
\begin{align} \label{DressedDelta}
 - \hh \Delta^{j^\prime}_j \equiv & -\hh \braD{j} \Delta \sigma_x \ketU{j^\prime} \nonumber \\
     =& -\hh \Delta \, [\sign{j^\prime-j}]^{|j^\prime-j|} \DXi{|j^\prime -j|}{\Min{j,j^\prime}} (\alpha),
\end{align} 
with 
 \begin{equation}
 \DXi{l}{j} (\alpha)=  \alpha^{l/2} \sqrt{\frac{j!}{(j+l)!}}  \mathsf{L}^{l}_j (\alpha) \rme^{-\frac{\alpha}{2}},
\end{equation} 
 and $\alpha = (2 g / \Omega)^2$. This dressing by Laguerre polynomials becomes in the high photon limit, $j \rightarrow \infty$, and for finite $l$ a dressing by Bessel functions, just like in the case of a classically driven TLS \cite{Grifoni1998, Wilson2007, Wilson2010, Hausinger2010}. \newline
For $\Delta =0$ and $\eps = l \Omega$, the unperturbed eigenstates $\ketD{j}$ and $\ketU{j+l}$ are degenerate, so that we can identify a twofold degenerate subspace in the complete Hilbert space of the problem \footnote{Notice, that for $l >0$ the first $l$ spin-up states have no degenerate partner, while for  $l<0$ the first $l$ spin-down states are unpaired.}. By using VVP \cite{VanVleck1929}, we can determine an effective  Hamiltonian $H_\text{eff}  = \exp(\rmi S) H \exp(-\rmi S)$ for the perturbed system consisting of 2 by 2 blocks of the shape
\begin{equation} \label{BlockHeff}
	\left( \begin{array}{cc}
	E_{\downarrow,j}^0 + \frac{\hbar}{4}\eps^{(2)}_{\downarrow,j} &-\hh \Delta_j^{j+l} \\
	-\hh \Delta_j^{j+l} & E_{\uparrow, j+l}^0 - \frac{\hbar}{4} \eps^{(2)}_{\uparrow, j+l}
	\end{array}
 \right),
\end{equation}
where we calculate the transformation matrix $S$ to second order in $\Delta$ \footnote{In \cite{Hausinger2010} similar calculations have been performed for a TLS coupled to a classical oscillator. They can be easily generalized to the quantized case.} and define the diagonal corrections as 
\begin{equation}
  \eps^{(2)}_{\downarrow/\uparrow,j} = \sum_{\substack{k=-j \\k \neq \pm l}}^\infty \frac{ \left(\Delta_{j}^{k+j}\right)^2}{\eps\mp k \Omega}.
\end{equation} 
Notice that for zero bias, $\eps=0$, the degenerate subspace consists of oscillator states with equal quantum number $j$. If one neglects the second-order corrections $\eps^{(2)}$ the effective Hamiltonian reduces to the one obtained within the ``adiabatic approximation'' in  \cite[see Eq. (9) there]{Irish2005}. Thus, our approach automatically also includes the adiabatic approximation. In \cite{Irish2005} only the zero bias case is considered; here we extend the adiabatic approximation to finite bias disregarding the second order correction $\eps^{(2)}$ in Eq. (\ref{BlockHeff}).
 In \cite{Ashhab2010}, a finite bias $\eps$ is considered in the parameter regime where eigenstates with same oscillator quanta $j$ remain quasidegenerate, so that the tunneling matrix element of a subspace remains dressed by a $L_j^0$ Laguerre polynomial. This is a valid approximation in the case that $\Omega \gg \Delta_\text{b}$. On the contrary, when $\eps \gtrsim \Omega$ and therefore also $\Delta_\text{b} \gtrsim \Omega$, a dressing by higher-order Laguerre polynomials occurs even in first order in $\Delta$.
The eigenenergies of Eq. (\ref{BlockHeff}) are
\begin{equation}  \label{VVEnergies}
 E_{\mp,j} = \hbar \left[ \left(j+ \frac{l}{2}\right) \Omega - \frac{g^2}{\Omega} + \frac{1}{8} \left(\eps_{\downarrow,j}^{(2)}-\eps_{\uparrow,j+l}^{(2)}\right) \mp \frac{1}{2} \Omega_j^l\right]
\end{equation} 
with the \textit{dressed oscillation frequency}
\begin{equation} \label{DressedOsc}
 \Omega_j^l =\sqrt{\left[ \eps - l \Omega + \frac{1}{4} \left(\eps_{\downarrow,j}^{(2)}+\eps_{\uparrow,j+l}^{(2)}\right) \right]^2+ \left(\Delta_j^{j+l}\right)^2}.
\end{equation} 
Notice that the quantum number $j$ corresponds to a \textit{mixture} of the oscillator levels $j$ and $l$. Only for $\eps =0$ this mixing vanishes.
We obtain the eigenstates of $H$ by $\ket{\Phi_{\pm,j}} = \exp(-\rmi S) \ket{\Phi_{\pm,j}^{(0)}}$ with the  eigenstates of (\ref{BlockHeff}) given by
\begin{align}
  \ket{\Phi_{-,j}^{(0)}} &= -\sinmixa \ketD{j} -\sign{\Delta_j^{j+l}} \cosmixa \ketU{j+l}, \label{FiMineff}\\
 \ket{\Phi_{+,j}^{(0)}} &= \cosmixa \ketD{j} -\sign{\Delta_j^{j+l}} \sinmixa \ketU{j+l}, \label{FiPluseff}
\end{align} 
and the mixing angle
\begin{equation}
 \tan \Mixangle = \frac{|\Delta_j^{j+l}|}{\eps - l \Omega + \frac{1}{4} \left(\eps_{\downarrow,j}^{(2)}+\eps_{\uparrow,j+l}^{(2)}\right)}
\end{equation} 
for $0< \Mixangle \leq \pi$. In Appendix \ref{App::Eigenstates}, the transformation is calculated to second order in $\Delta$ and applied to the effective states. By this we have all information we need to calculate the dynamics of the qubit-oscillator system.\newline
Van Vleck perturbation theory yields good approximate results as long as the matrix elements connecting different non-degenerate subspaces with each other are much smaller then the energetical distance between those subspaces \cite{CohenTannoudji2004}. In our case this means
\begin{equation} \label{ValidVV}
  \left|\frac{1}{2} \Delta_j^{j+k}\right| \ll |\eps - k \Omega| \quad \forall \quad k \neq l. 
\end{equation} 
We will discuss the validity of our approach for the different cases below.

\section{Energy spectrum in the ultrastrong coupling regime} \label{Sec::EnergySpec}
In this section, we examine the energy spectrum of the qubit-oscillator system as  obtained from Eq. (\ref{VVEnergies}) and compare it to results found by exact numerical diagonalization. We check its robustness for variable coupling strength $g$ and detuning $\delta = \Delta_\text{b} - \Omega$ between the qubit energy splitting and oscillator frequency.

\subsection{Zero static bias $\eps =0$} \label{Sec::EnergySpecZeroBias}
First, we concentrate on the regime of zero static bias. This is the usual case in cavity QED, where the JCM is applied.  The JCM  is known to work well for weak qubit-oscillator  coupling ($g/\Omega \ll 1$) and small detuning between the two devices. As already predicted in \cite{Hausinger2008}, higher-order corrections have to be taken into account for stronger coupling. For the case of ultrastrong coupling, we will find that the situation changes dramatically. The energies predicted by the JCM read
\begin{equation} \label{EJCM}
 E_{2j+1,2j+2}^\text{JCM} = \hbar \biggl[\left(j+\frac{1}{2}\right) \Omega \mp \frac{1}{2} \sqrt{(\Delta - \Omega)^2 + 4(j+1)g^2}\biggr]
\end{equation}
with the ground state energy  $E_0^\text{JCM} = - \hbar \Delta/2$. Equation (\ref{VVEnergies}) for the Van Vleck eigenenergies perturbative in $\Delta$, simplifies further for $\eps =0$:
\begin{equation} \label{VVEnE0}
 E_{\mp, j} = \hbar \biggl[ j \Omega - \frac{g^2}{\Omega} - \frac{1}{4} \sum_{\substack{k=-j \\ k \neq 0}}^\infty \frac{(\Delta_j^{k+j})^2}{k \Omega} \mp \frac{1}{2} |\Delta \mathsf{L}_j^0 (\alpha) \rme^{-\alpha/2} | \biggr].
\end{equation} 
The semi-infinite sum in the above expression converges, and we show in Appendix \ref{App::VVEnE0} analytical expressions for the first four energy levels. Furthermore, we can compare our results to the generalized rotating-wave approximation (GRWA) \cite{Irish2007}. In this approach, the total Hamiltonian Eq. (\ref{HTot}) is expressed in the displaced basis states of the adiabatic approximation. It is then in this representation, that the rotating-wave approximation is performed and counter-rotating terms are neglected. Thus, the GRWA uses the advantages of the adiabatic approximation, namely its ability to go to strong coupling strengths and to treat detuned systems, and  also gives reliable results in the weak coupling regime of the JCM. A derivation of the GRWA eigenenergies can be found in Appendix \ref{App::GRWA}.

\subsubsection{Energy levels against detuning}
\begin{figure}
\centering
 \includegraphics[width=8.6cm]{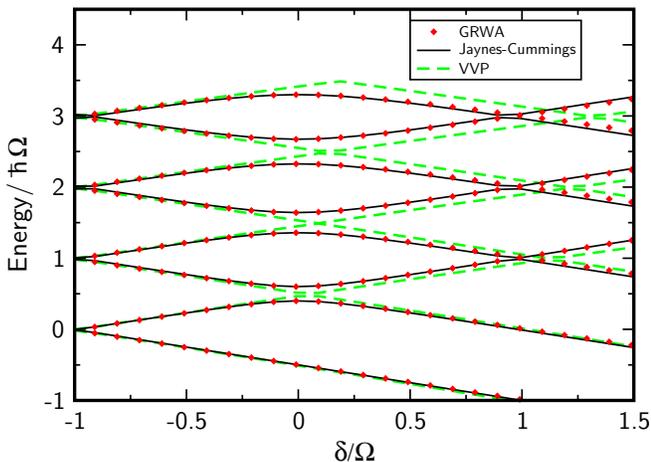}
 \caption{Energy levels against detuning $\delta = \Delta - \Omega$ for $\eps/\Omega=0$, $g/\Omega=0.1$. Our VVP solution is compared to the GRWA and the JCM. The latter two agree  well with numerical calculations for the whole detuning range (not shown), while VVP yields only reliable results for negative detuning, $\Delta < \Omega$. \label{Fig::EnergyVSDet_e=0_g=0.1}}
\end{figure}

\begin{figure}
 \includegraphics[width=8.6cm]{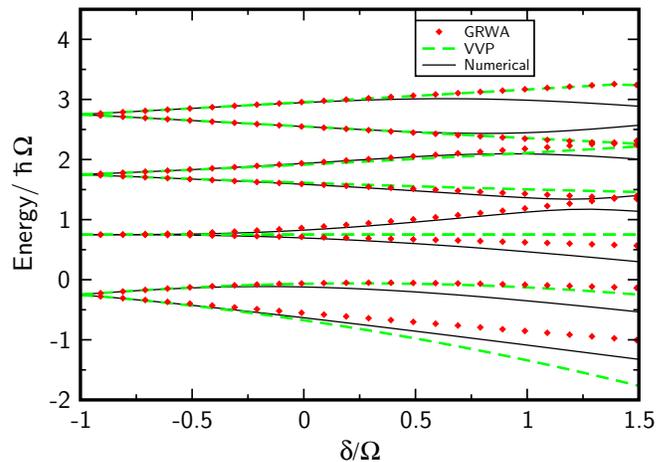}
 \caption{Energy levels against detuning $\delta = \Delta - \Omega$ for $\eps/\Omega=0$, $g/\Omega=0.5$. The JCM fails already completely for such a coupling strength (not shown). We compare VVP and  the GRWA against numerical calculations. Both agree well with the numerics for negative detuning and even at resonance. For stronger positive detuning they both fail and strongest deviations can be seen for the lower energy levels. \label{Fig::EnergyVSDet_e=0_g=0.5}}
\end{figure}

\begin{figure}
 \includegraphics[width=8.6cm]{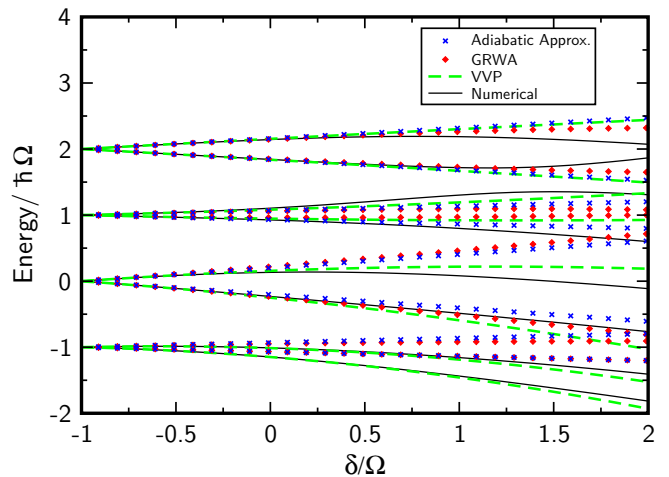}
 \caption{Energy levels against detuning $\delta = \Delta - \Omega$ for $\eps/\Omega=0$, $g/\Omega=1.0$.  We compare VVP, the adiabatic approximation and GRWA against a numerical calculation. For a negative detuning all three approaches agree very well with the exact numerics. However, for zero and positive detuning deviations occur.  In particular, the ground level and the first excited level are not described correctly by the adiabatic approximation and the GRWA for strong positive detuning, while VVP yields good results.  \label{Fig::EnergyVSDet_e=0_g=1.0}}
\end{figure}

\begin{figure}
 \includegraphics[width=8.6cm]{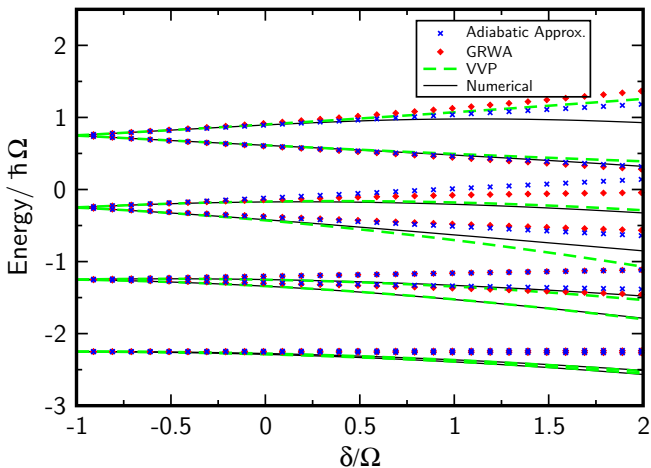}
 \caption{Energy levels against detuning. Same as in Fig. \ref{Fig::EnergyVSDet_e=0_g=1.0}, but for a coupling strength of $g/\Omega =1.5$. Adiabatic approximation and GRWA fail for  positive detuning, while VVP  gives the first four energy levels correctly even up to a detuning of $\delta/\Omega =2.0$. And also for the higher energy levels it yields good results beyond the resonant case.  \label{Fig::EnergyVSDet_e=0_g=1.5}}
\end{figure}

In Figs. \ref{Fig::EnergyVSDet_e=0_g=0.1} -- \ref{Fig::EnergyVSDet_e=0_g=1.5} we examine the energy levels  against the qubit-oscillator detuning $\delta = \Delta - \Omega$ at fixed couplings, $g/\Omega=0.1$, $0.5$, $1.0$ and $1.5$, respectively.\newline
For a weak coupling of $g/\Omega=0.1$, we compare VVP to the GRWA and the JCM. Both are known to work well in this regime.
We find that VVP gives only valid results for negative detuning, $\Delta < \Omega$. This was expected as it relies on a perturbative approach in $\Delta$, and we  know already from the adiabatic approximation that it fails for $\Delta \gtrsim \Omega$ and simultaneously small $g/\Omega$. In this regime of weak coupling, the JCM or GRWA are clearly preferable to our method.
\newline
For an intermediate coupling strength, the same discussion is presented in Fig. \ref{Fig::EnergyVSDet_e=0_g=0.5}. We do not show the Jaynes-Cummings energy levels in this regime anymore, because they fail completely to return the correct energy spectrum. Instead we compare to a numerical diagonalization of the Hamiltonian. Van Vleck perturbation theory and the GRWA yield good results for negative detuning $\delta < 0$, but also at resonance, $\Delta = \Omega$, they agree relatively well with the numerics. At positive detuning both deviate strongly from the exact solution. 
\newline
With a coupling strength of $g/\Omega = 1.0$ in Fig. \ref{Fig::EnergyVSDet_e=0_g=1.0}, we are already deep in the ultrastrong coupling regime. Those high values have not been observed experimentally yet. They are, however, predicted to be realizable \cite{Devoret2007}.  For negative detuning, GRWA and VVP show a good agreement with the numerics. However, approaching zero detuning or going beyond to positive one, the GRWA fails in particular for the two lowest states, which will turn out to be important for the calculations of the dynamics. In order to explain this failure, we also show in Fig. \ref{Fig::EnergyVSDet_e=0_g=1.0} the adiabatic approximation. As pointed out, the GRWA is a combination of the ordinary RWA, and thus works well for weak coupling, and of the adiabatic approximation, which works very well for strong negative detuning, $\Omega \gg \Delta$, for all values of the coupling. At resonance or at positive detuning, the adiabatic approximation shows deviations from the exact solution for a coupling strength $g/\Omega =1.0$. This coupling strength is, however, already too strong to be treated correctly by the RWA. Thus, we are in a kind of intermediate regime, which is also not covered by the GRWA, but can be important in experimental applications. On the contrary, VVP shows an exact agreement with the numerical data for negative detuning and even up to exact resonance. Only for positive detuning, deviations start to occur.\newline
This becomes even more prominent for stronger coupling strengths, like $g/\Omega =1.5$ in Fig. \ref{Fig::EnergyVSDet_e=0_g=1.5}. 
While the adiabatic approximation and also the GRWA fail for positive detuning, VVP agrees surprisingly well with the numerical results up to $\delta =2.0$ for the first four energy levels; i.e., we have $\Delta/\Omega =3.0$. Also for the higher levels we still find a good agreement for not too strong positive detuning. This improvement is due to the fact that VVP also takes into account connections between non-degenerate subspaces and therefore higher-order corrections in the dressed tunneling matrix element. 

\subsubsection{Energy levels against coupling strength}
\begin{figure}
 \includegraphics[width=8.6 cm]{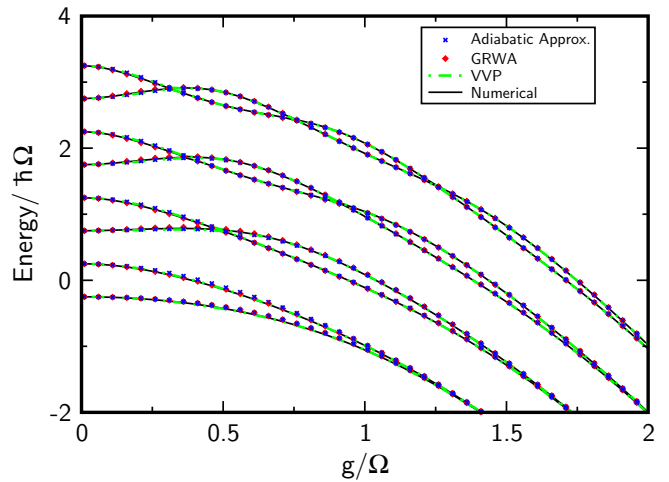}
 \caption{Energy levels against coupling strength $g/\Omega$ for negative detuning $(\delta/\Omega =-0.5)$. Numerical results are compared with the adiabatic approximation, GRWA and VVP. All three approaches show only slight deviations.  \label{Fig::EnergyVSg_e=0_W=1_D=0.5}}
\end{figure}

\begin{figure}
 \includegraphics[width=8.6cm]{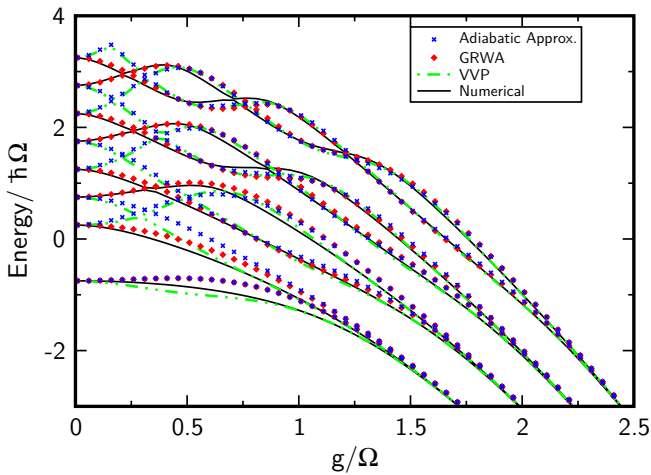}
 \caption{Energy levels against coupling strength for positive detuning $(\delta/\Omega = 0.5)$. For coupling strengths with $g/\Omega \gtrsim 0.75$, VVP exhibits the best agreement with numerical results, while for smaller coupling and higher energy levels, the GRWA should be used.   \label{Fig::EnergyVSg_e=0_W=1_D=1.5}}
\end{figure}

In Figs. \ref{Fig::EnergyVSg_e=0_W=1_D=0.5} -- \ref{Fig::EnergyVSg_e=0_W=1_D=1.5} we investigate now the first eight energy levels against the coupling strength $g/\Omega$ for three different values of the detuning.
\newline
All three approaches, the adiabatic approximation, the GRWA and VVP, show very good agreement with the numerical results for the whole range of $g/\Omega$ for negative detuning $\delta/\Omega = -0.5$  shown in Fig. \ref{Fig::EnergyVSg_e=0_W=1_D=0.5}.\newline
At resonance, $\Delta / \Omega =1.0$, in Fig. \ref{Fig::EnergyVSg_e=0_W=1_D=1}, we have to distinguish between different parameter regimes:
\begin{figure}
 \includegraphics[width=8.6cm]{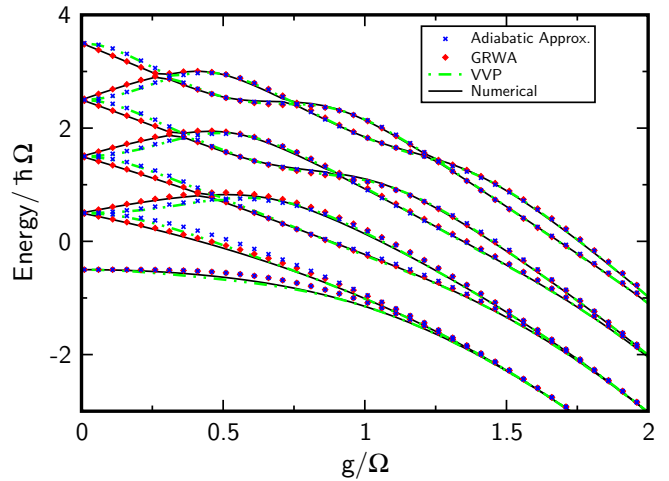}
 \caption{Energy levels against coupling strength at resonance $(\delta /\Omega =0)$. For small coupling strength, the adiabatic approximation and VVP show small deviations from the correct values (see especially the higher energy levels). The GRWA works well in this regime. For stronger coupling strength, all three approaches agree well with the numerical results.   \label{Fig::EnergyVSg_e=0_W=1_D=1}}
\end{figure}
For smaller values of the coupling, $g/\Omega \lesssim 0.5$, the adiabatic approximation and VVP show deviations from the numerical results apart from the ground level, as they do not take into account correctly the zero coupling resonance \cite{Irish2007}, while the GRWA on the other hand works well. For higher coupling strengths on the other hand, VVP exhibits a slight improvement  to the GRWA and the adiabatic approximation for the first two energy levels, as  could already be seen from Figs. \ref{Fig::EnergyVSDet_e=0_g=1.0} and \ref{Fig::EnergyVSDet_e=0_g=1.5}.  \newline
This improvement becomes more evident for stronger positive detuning, $\delta/\Omega =0.5$, as shown in Fig. \ref{Fig::EnergyVSg_e=0_W=1_D=1.5}.
Considering the lowest two energy levels, VVP agrees well with the numerical results for $g/\Omega \gtrsim 0.75$, while the adiabatic approximation and GRWA strongly deviate from the numerical results. For higher levels also the latter two are closer to the numerics. However, for weaker couplings the results from all three approaches are not very satisfying even for the lower energy levels, and the adiabatic approximation and VVP predict unphysical crossings, while the GRWA at least yields the correct weak coupling limit.\newline
Plotted against the coupling strength the energy levels exhibit some peculiarities.  Most interesting is the finding that for strong coupling two adjacent energy levels become degenerate, so that coherent oscillations between them become completely suppressed. We can understand that by considering expression (\ref{VVEnE0}), where we find that two energy levels with the same index $j$ differ only in the sign of the dressed oscillation frequency, which vanishes for large $g$. For the higher energy levels, degeneracies also occur for lower $g/\Omega$ values, happening at the zeros of the Laguerre polynomials. These phenomena are discussed in more detail in \cite{Irish2005, Irish2007, Ashhab2010}, and we come back to them when presenting the dynamics.

\subsubsection{Validity regimes}
\begin{figure}
 \includegraphics[width=8.6cm]{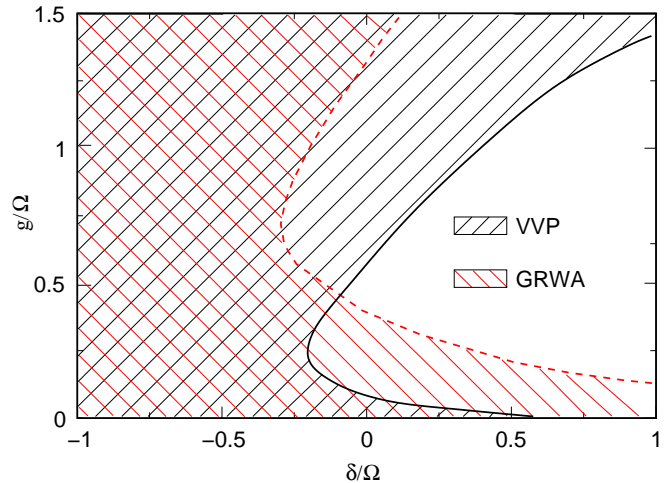}
 \caption{Sketch of the validity regime of VVP and GRWA for $\eps=0$. The GRWA is perferable to VVP at weak coupling, in particular close to resonance and positive detuning. On the contrary, VVP works better at strong coupling strengths. \label{Fig::CompVVPGRWA}}
\end{figure}

To summarize this section we give a comparison between VVP and the GRWA. We do not discuss the adiabatic approximation and the JCM as they are included in VVP and the GRWA, respectively. Further, we want to emphasize that Fig. \ref{Fig::CompVVPGRWA} only represents a qualitative sketch; the detailed behavior is more complicated:  
the validity regime of the different approaches is crucially dependent on the error one allows compared to numerical solutions. Furthermore, the number of energy levels taken into account plays a role. For instance, in Fig. \ref{Fig::EnergyVSg_e=0_W=1_D=1.5} VVP agrees very well with the numerics for the lowest two energy levels and $g/\Omega \approx 0.75$, but shows already stronger deviations for the 5th and 6th level. In Fig. \ref{Fig::CompVVPGRWA} we took the first eight levels into account. 
In order to understand the validity regime of VVP we consider Eq. (\ref{ValidVV}) for $\eps =0$. In this special case it becomes
\begin{equation}
 \left|\frac{1}{2} \Delta_j^{j+k}\right| \ll |k \Omega| \quad \forall \quad k \neq 0.
\end{equation} 
From the definition of the dressed tunneling matrix element $\Delta_j^{j+k}$, Eq. (\ref{DressedDelta}), we see that for small $\Delta /\Omega$ -- i.e., for negative detuning -- this condition is fullfilled even for weak coupling. However, for $\Delta  \gtrsim\Omega$ \textit{and} weak coupling, the above condition does not hold anymore. On the other hand, by increasing the coupling strength VVP becomes even valid at strong positive detuning since the dressed tunneling matrix elements are exponentially suppressed. The GRWA is valid for positive detuning also in the case of weak coupling. For intermediate coupling $0.5 \lesssim g/\Omega \lesssim 1.0$ it fails for zero or positive detuning, while increasing the coupling strength further yields  an improvement in this regime. This last tendency has the same origin as in case of VVP, namely that the neglected tunneling matrix elements get suppressed. As, however, the GRWA considers these matrix elements only to first order, the improvement is not as good as for VVP.

\subsection{Finite static bias $\eps \neq0$}
\begin{figure}
 \includegraphics[width=8.6cm]{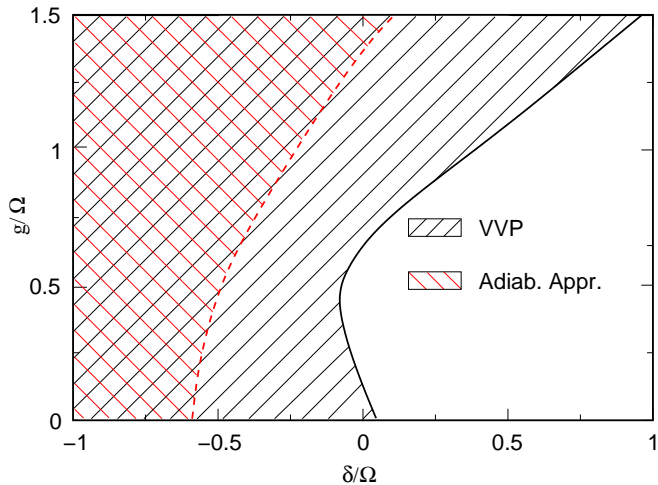}
 \caption{Sketch of the validity regime of VVP and of the adiabatic approximation for $\eps=1.0$. For positive detuning and simultaneously weak coupling both approaches fail. For stronger coupling VVP yields an improvement to the adiabatic approximation. \label{Fig::CompVVPAdiab}}
\end{figure}

\begin{figure}
 \includegraphics[width=8.6cm]{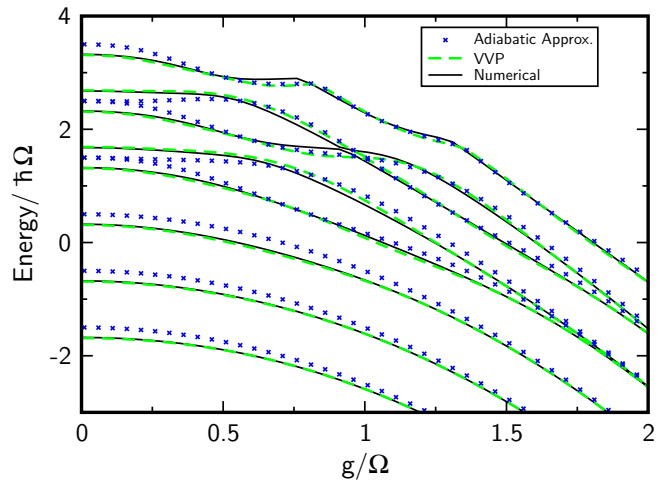}
 \caption{Energy levels against coupling $g/\Omega$ for $\eps/\Omega =3.0$ and $\Delta/\Omega = 1.5$. The three lowest energy levels have no degenerate partner. Despite the high value of $\Delta$, VVP still gives reliable results, while the adiabatic approximation differs from the numerical values even for the low energy levels. \label{Fig::EnergyVSGe3_D=1.5} }
\end{figure}

In this section we  discuss the energy spectrum for the case of finite static bias. We compare our VVP calculation to exact numerical diagonalization. We further show in certain cases calculations disregarding connections between the different manifolds, that is second-order corrections in $\Delta$, which is the natural extension of the adiabatic approximation to finite bias. We do not compare to the GRWA, as it exists so far just for the zero bias case.
\begin{figure}
 \includegraphics[width=8.6cm]{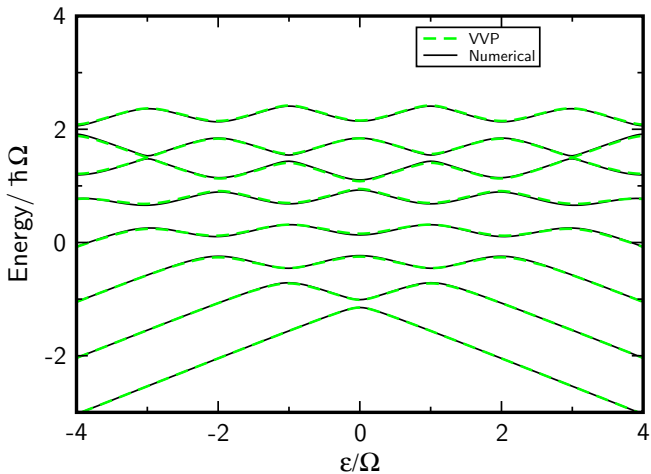}
 \caption{Energy levels against static bias $\eps$ for $g/\Omega=1.0$ at resonance $\Delta/\Omega =1.0$. Van Vleck perturbation theory is compared to a numerical diagonalization of the Hamiltonian. \label{Fig::EnergyVSe_D=1}}
\end{figure}
 To start, we show in Fig. \ref{Fig::EnergyVSe_D=1} the energy levels against the static bias for a coupling strength of $g/\Omega =1.0$ and no  detuning in the zero bias case ($\Delta = \Omega$). For such a coupling strength, we find a very good agreement between our VVP calculations and numerically obtained results. Most remarkably, this agreement holds even away from the resonant points, $\eps = l \Omega$, for which our approximation has been performed. We also checked the effect on the spectrum when neglecting the second-order corrections in $\Delta$. The qualitative behavior remains the same; however, quantitative deviations occur (not shown in Fig. \ref{Fig::EnergyVSe_D=1}). For negative detuning, $\Delta < \Omega$, the agreement between analytical and numerical results is even enhanced, while for positive detuning up to $\Delta/\Omega = 1.5$ only slight deviations occur. The accuracy of VVP diminishes entering the weak coupling regime, as we could already observe for the zero static bias case and we will show in the following. Before, we want to consider some general features of the spectrum at nonzero static bias. We already pointed out while identifying the degenerate subspaces in Eq. (\ref{BlockHeff}) that for $\eps = l \Omega$ with $l \neq 0$ certain unperturbed energy levels have no degenerate partner. Without loss of generality, we assume $l>0$, that means that the first $l$ energy levels corresponding to a spin-up state have no degenerate partner and their energy is simply given by $E_{\uparrow, j}^0 - \frac{\hbar}{4} \eps^{(2)}_{\uparrow, j}$ with $j = 0,1,2,... l-1$. Of course, also the corresponding effective eigenstates are simply $\ketU{j}$, and we cannot observe avoided crossings or a superposition of states. For instance, in Fig. \ref{Fig::EnergyVSe_D=1} at $\eps/\Omega =1$, we observe the lowest energy level being without partner, while the higher ones form avoided crossings with the adjacent level. For $\eps/\Omega =2$, the two lowest levels are ``free'', etc. \newline
\begin{figure}
 \includegraphics[width=8.6cm]{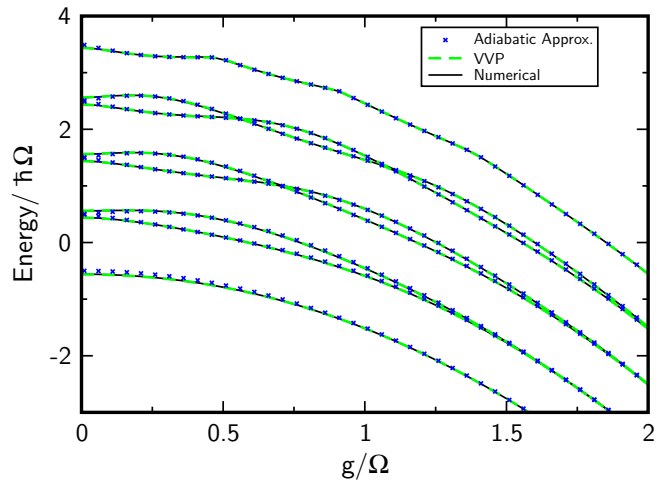}
 \caption{Energy levels against coupling $g/\Omega$ for $\eps/\Omega =1.0$ and $\Delta/\Omega = 0.5$. The adiabatic approximation and VVP agree almost perfectly with numerical results. Slight deviations can be seen for the adiabatic approximation at $g/\Omega \to 0$. \label{Fig::EnergyVSGe1_D=0.5}}
\end{figure}
\begin{figure}
 \includegraphics[width=8.6cm]{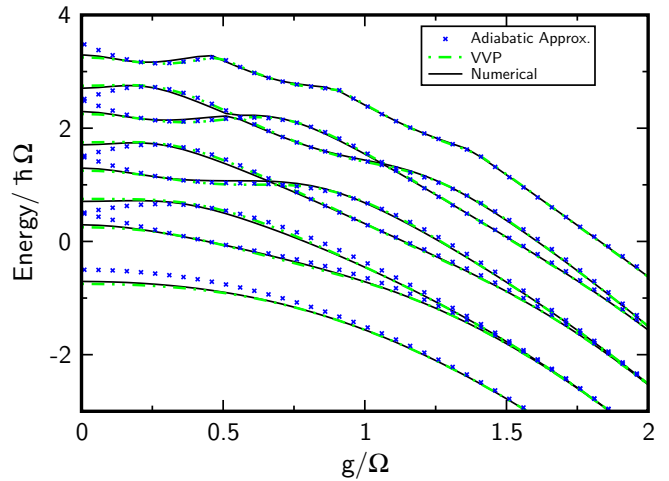}
 \caption{ Energy levels against coupling $g/\Omega$ for $\eps/\Omega =1.0$ and $\Delta/\Omega = 1.0$. Van Vleck perturbation theory is still valid compared to numerical results, while the adiabatic approximation fails specifically for weak coupling strengths. \label{Fig::EnergyVSGe1_D=1}}
\end{figure}
\begin{figure}
 \includegraphics[width=8.6cm]{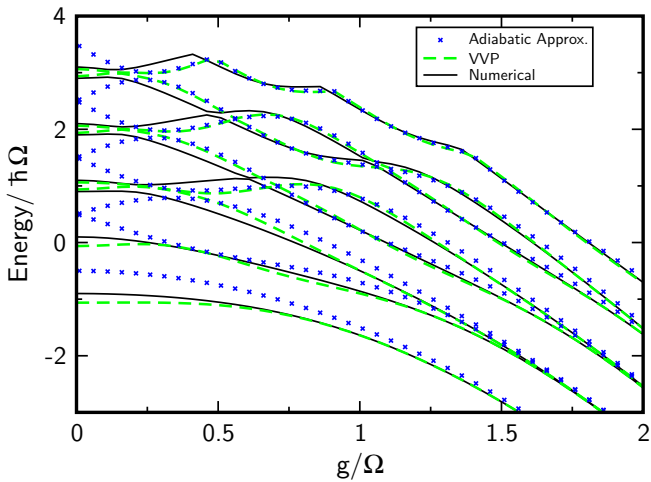}
 \caption{ Energy levels against coupling $g/\Omega$ for $\eps/\Omega =1.0$ and $\Delta/\Omega = 1.5$. In this regime, also VVP shows deviations from the numerical results for $g/\Omega \lesssim 0.75$ especially for the higher energy levels. It agrees well for stronger coupling. \label{Fig::EnergyVSGe1_D=1.5} }
\end{figure}
In Figures  \ref{Fig::EnergyVSGe1_D=0.5}, \ref{Fig::EnergyVSGe1_D=1} and \ref{Fig::EnergyVSGe1_D=1.5}, we present the dependence of the energy spectrum on the coupling strength $g/\Omega$ for the case of $\eps/\Omega =1.0$ and $\Delta/\Omega = 0.5$, $\Delta/\Omega = 1.0$ and $\Delta /\Omega =1.5$, respectively.
Just like in the zero static bias case, VVP yields best results for $\Delta /\Omega < 1$, because there the condition for a perturbative approach is most satisfied. Also, the extended adiabatic approach yields very convincing results, only for $g/\Omega \to 0$ one can notice slight deviations. For $\Delta /\Omega = 1.0$ in Fig. \ref{Fig::EnergyVSGe1_D=1}, VVP still shows almost exact agreement with the numerical results, whereas the adiabatic approximation fails for weak coupling. This failure of the latter becomes more evident going to positive detuning like $\Delta/\Omega=1.5$ in Fig. \ref{Fig::EnergyVSGe1_D=1.5}. But there also the VVP exhibits strong deviations for coupling strengths $g/\Omega \lesssim 0.75$.\newline
Figure \ref{Fig::CompVVPAdiab} summarizes these observations in a qualitative sketch of the validity regimes. Thereby VVP excels the adiabatic approximation as it considers also second-order corrections in the matrix elements connecting different doublets.
\newline
We also tested for static bias values being no multiples of $\Omega$ and found a confirmation of the above findings. For stronger static bias, VVP describes the lower energy levels even better for positive detuning, see, e.g., the case $\eps/\Omega = 3.0$ in Fig. \ref{Fig::EnergyVSGe3_D=1.5}. Here, the three lowest energy levels are without degenerate partner and therefore can be described by the corrected unperturbed energy. The influence of the mixing to other energy levels is less strong.

\section{Dynamics of the qubit in the ultrastrong coupling regime} \label{Sec::Dynamics}
\begin{figure}
 \includegraphics[width=8.6cm]{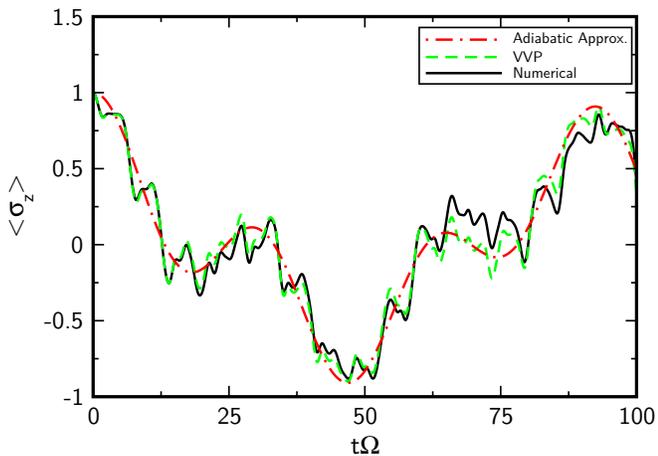}
 \caption{Population difference for zero static bias. Further parameters are $\Delta/\Omega = 0.5$, $\hbar \beta \Omega = 10$ and $g/\Omega =1.0$. The adiabatic approximation and VVP are compared to numerical results.  The first one only covers the longscale dynamics, while VVP also returns the fast oscillations. With increasing time small differences between numerical results and VVP become more pronounced. \label{Fig::P_e=0_D=0.5_g=1.0}}
\end{figure}
\begin{figure*}
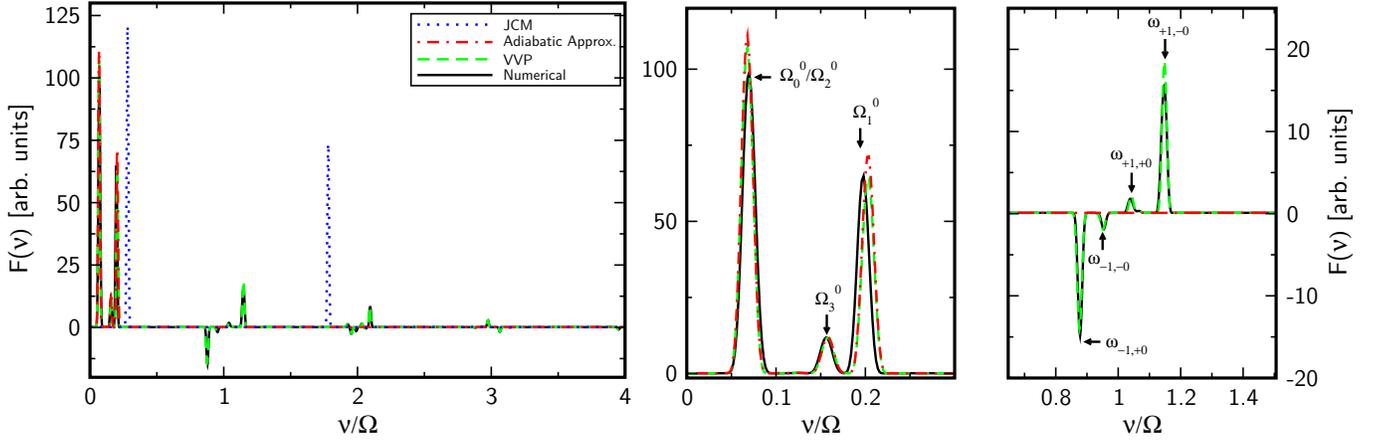

 \includegraphics[width=8.3cm]{Figures/Fig_16a_Revised.eps}
\includegraphics[width=9.5cm]{Figures/Fig_16b_Revised.eps}
 \caption{Fourier transform of the population difference in Fig. \ref{Fig::P_e=0_D=0.5_g=1.0}. The left-hand graph shows the whole frequency range. The lowest frequency peaks originate from transitions between levels of a degenerate subspace and are determined through the dressed oscillation frequency $\Omega_j^0$. Numerical calculations and VVP predict group of peaks located around $\nu/\Omega = 0,1.0,2.0,3.0$. The first group at $\nu/\Omega =0$ is shown in the middle graph. One can identify frequencies $\Omega_0^0$ and $\Omega_2^0$, which fall together, and $\Omega_1^0$. The small peak comes from the frequency $\Omega_3^0$.  This first group of peaks is also covered by the adiabatic approximation. The other groups come from transitions between different manifolds. The adiabatic approximation does not take them into account, while VVP does. A blow-up of the peaks coming from transitions between neighboring manifolds is given in the right-hand graph. In the left-hand graph additionally the Jaynes-Cummings peaks are shown, which, however, fail completely. \label{Fig::F_e=0_D=0.5_g=1.0}}
\end{figure*}
We are interested in determining the population difference between the two qubit states; i.e., we calculate
\begin{equation}
  \langle \sigma_z(t) \rangle = \text{Tr}_\text{TLS}\{ \sigma_z \rho_\text{red}(t) \} = 2 \bra{\uparrow} \rho_\text{red}(t) \ket{\uparrow} - 1,
\end{equation} 
where $\rho_\text{red}(t)$ is obtained after tracing out the oscillator degrees of freedom from the qubit-oscillator density operator $\rho$. The matrix elements of the latter  read in the system's energy eigenbasis $\{ \ket{\Phi_{\alpha=\{\pm, j\}}} \}$
\begin{equation}
 \rho_{\alpha \gamma} (t)=\bra{\Phi_\alpha} \rho(t) \ket{\Phi_\gamma} = \rho_{\alpha \gamma}(0) \rme^{-\rmi \omega_{\alpha \gamma} t}.
\end{equation}   
As starting conditions, we assume the qubit and the oscillator to be uncoupled for $t<0$, and the first to be prepared in the spin-up state, with the oscillator being in thermal equilibrium:
\begin{equation}
 \rho(0) = \ket{\uparrow}\bra{\uparrow}  \otimes \sum_j \frac{1}{Z} \rme^{ - \hbar \beta j \Omega} \ket{j}\bra{j},
\end{equation} 
where $Z$ is the partition function of the harmonic oscillator and $\beta$ the inverse temperature. In the following, we will assume $\hbar \beta \Omega = 10$, which corresponds for oscillator frequencies in the GHz regime to experiments performed at several mK. At those low temperatures, mainly the lower oscillator energy levels are of importance. The dynamics for higher oscillator occupation numbers at zero static bias has been investigated in \cite{Irish2005}.\newline
The transition frequencies are defined as $\omega_{\alpha \gamma} =(E_\alpha - E_\gamma)/\hbar$, where $E_\alpha$ stands either for $E_{\mp, j}$ in case of two-fold degenerate subspaces or $E_{\uparrow/\downarrow,j}^0 \mp \frac{\hbar}{4} \eps^{(2)}_{\uparrow/\downarrow,j}$ for one-dimensional subspaces. We further can distinguish between two different timescales: large oscillatory contributions are resulting from different oscillator quanta $j$, while the difference in dressed oscillation frequencies $\Omega_j^l$ acts on a much longer timescale and its contribution vanishes for large coupling strengths $g/\Omega$.\newline
In the following subsections we will investigate the dynamics for the unbiased and biased case. Again, we will compare exact numerical results 
to VVP and the adiabatic approximation. Apart from the energy levels, also the eigenstates become now of importance. In particular, we will find that away from the condition $\eps = l \Omega$, the higher-order corrections are crucial to give the correct dynamics.

\subsection{Dynamics for zero static bias $\eps =0$}
\begin{figure}
 \includegraphics[width=8.6cm]{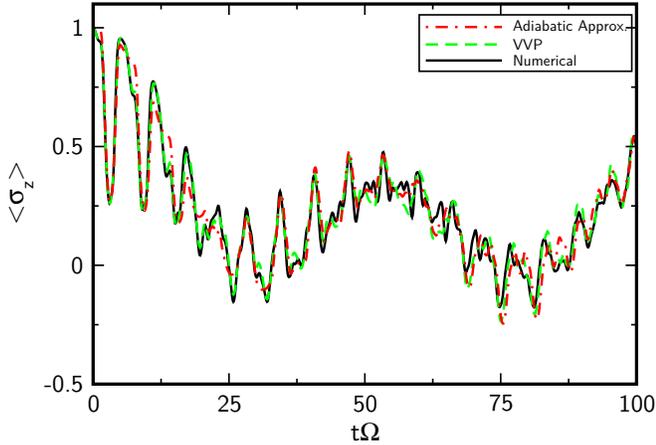}
 \caption{Population difference for zero static bias. Same parameters as in Fig. \ref{Fig::P_e=0_D=0.5_g=1.0} but for a coupling strength of $g/\Omega =2.0$. Both the adiabatic approximation and VVP agree well with the numerics, but show slight dephasing on a longer timescale.\label{Fig::P_e=0_D=0.5_g=2.0} }
\end{figure}
\begin{figure*}
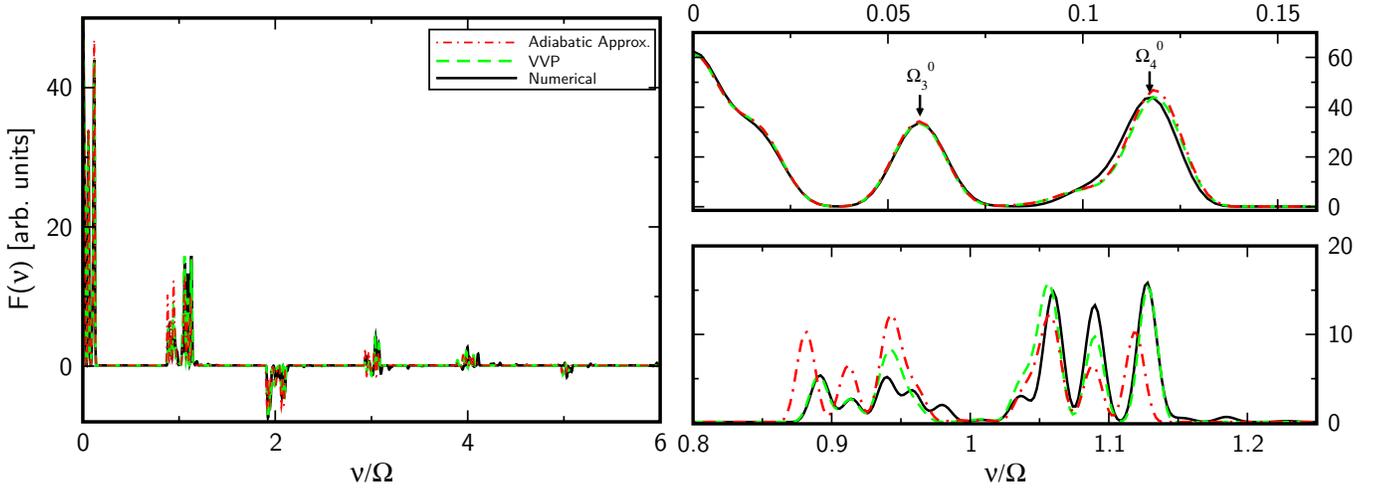

 \includegraphics[width=8.8cm]{Figures/Fig_18a_Revised.eps}
\includegraphics[width=9.0cm]{Figures/Fig_18b_Revised.eps}
 \caption{Fourier spectrum of the population difference in Fig. \ref{Fig::P_e=0_D=0.5_g=2.0}. In the left-hand graph a large frequency range is covered. Peaks are located around $\nu/\Omega = 0$, $1.0$, $2.0$, $3.0$ etc. Even the adiabatic approximation exhibits the higher frequencies. The upper right-hand graph shows the first group close to $\nu/\Omega =0$. The two main peaks come from $\Omega_3^0$ and $\Omega_4^0$ and higher degenerate manifolds. Frequencies from lower manifolds contribute to the peak at zero. The adiabatic approximation and VVP agree well with the numerics. The lower right-hand graph shows the second group of peaks around $\nu /\Omega = 1.0$. This group is also predicted by the adiabatic approximation and VVP, but they do not fully return the detailed structure of the numerics. Interestingly, there is no peak exactly at $\nu/\Omega =1.0$ indicating no nearest-neighbor transition between the low degenerate levels. \label{Fig::F_e=0_D=0.5_g=2.0}}
\end{figure*}
For zero static bias, we first examine a regime where we expect our approximation to work well. We thus consider a not too strong tunneling matrix element, $\Delta /\Omega =0.5$ and a coupling strength of $g/\Omega =1.0$. Figures \ref{Fig::P_e=0_D=0.5_g=1.0} and \ref{Fig::F_e=0_D=0.5_g=1.0} show the population difference $\langle \sigma_z(t) \rangle$ and its Fourier transform,
\begin{equation}
 F(\nu):= 2 \int_{0}^\infty dt \langle \sigma_z(t) \rangle \cos (\nu t),
\end{equation} 
respectively. Concerning the population difference, we see a relatively good agreement between the numerical calculation and VVP for short timescales. In particular, VVP also correctly returns the small overlaid oscillations. For longer timescales, the two curves get out of phase. The adiabatic approximation only can reproduce the coarse-grained dynamics. The fast oscillations are completely missed. To understand this better, we turn our attention to the Fourier transform in Fig. \ref{Fig::F_e=0_D=0.5_g=1.0}. There, we find several groups of frequencies located around $\nu/\Omega =0$, $\nu/\Omega =1.0$, $\nu/\Omega =2.0$ and $\nu/\Omega =3.0$. This can be explained by considering the transition frequencies in more detail. We have from Eq. (\ref{VVEnergies})
\begin{equation}
 \omega_{\mp k, \mp j}^{l} = \hbar [(k-j) \Omega + \zeta_{k,j}^{l} \pm \half (\Omega_j^l - \Omega_k^l)],
\end{equation} 
and
\begin{equation}
 \omega_{\mp k, \pm j}^{l} = \hbar [(k-j) \Omega + \zeta_{k,j}^{l} \mp \half (\Omega_j^l + \Omega_k^l)],
\end{equation} 
with $\zeta_{k,j}^{l} = \frac{1}{8} \left(\eps_{\downarrow, k}^{(2)}-\eps_{\downarrow, j}^{(2)}+\eps_{\uparrow, j+l}^{(2)}-\eps_{\uparrow, k+l}^{(2)}\right)$ being the second-order corrections. For zero bias, $\eps=0$, the index $l$ vanishes. The term $(k-j)\Omega$ determines to which group of peaks a frequency belongs and $\Omega_j^0$ its relative position within this group. The latter has $\Delta$ as an upper bound, so that the range over which the peaks are spread within a group increases with $\Delta$. The dynamics is dominated by the peaks belonging to transitions between the same subspace $k-j =0$, while the next group with $k-j=1$ yields already faster oscillations. To each group belong theoretically infinite many peaks. However, under the low temperature assumption only those with a small oscillator number play a role. For the used parameter regime, the adiabatic approximation does not take into account the connections between different manifolds. It therefore covers only the first group of peaks with $k-j=0$, providing the long-scale dynamics. For $\eps=0$, the dominating frequencies in this first group are given by  $\Omega_0^0 = |\Delta \rme^{-\alpha/2}|$, $\Omega_1^0 = |\Delta (1-\alpha)\rme^{-\alpha/2}|$ and $\Omega_2^0 = |\Delta \mathsf{L}_2^0(\alpha) \rme^{-\alpha/2}|$, where $\Omega_0^0$ and $\Omega_2^0$ coincide. A small peak at $\Omega_{3}^{0} = |\Delta \mathsf{L}_3^0(\alpha) \rme^{-\alpha/2}|$ can also be seen. Notice that for certain coupling strengths some peaks vanish; like, for example, choosing a coupling strength of $g/\Omega =0.5$ makes the peak at $\Omega_1^0$ vanish completely, independently of $\Delta$, and the $\Omega_0^0$ and $\Omega_2^0$ peaks split. The JCM yields two oscillation peaks determined by the Rabi splitting and fails completely to give the correct dynamics, see the left-hand graph in Fig. \ref{Fig::F_e=0_D=0.5_g=1.0}.\newline
Now, we proceed to an even stronger coupling, $g/\Omega =2.0$, where we also expect the adiabatic approximation to work better. From Fig. \ref{Fig::EnergyVSg_e=0_W=1_D=0.5} we noticed that at such a coupling strength the lowest energy levels are degenerate within a subspace. Only for oscillator numbers like $j=3$, we see that a small splitting arises. This splitting becomes larger for higher levels. Thus, only this and higher manifolds can give significant contributions to the long time dynamics; that is, they can yield low frequency peaks. Also the adiabatic approximation is expected to work better for such strong couplings \cite{Irish2005}.
And indeed by looking at Figs. \ref{Fig::P_e=0_D=0.5_g=2.0} and \ref{Fig::F_e=0_D=0.5_g=2.0}, we notice that both the adiabatic approximation and VVP agree quite well with the numerics. Especially the first group of Fourier peaks in Fig. \ref{Fig::F_e=0_D=0.5_g=2.0} is also covered almost correctly by the adiabatic approximation. The first manifolds we can identify with those peaks are the ones with $j=3$ and $j=4$. This is a clear indication that even at low temperatures higher oscillator quanta are involved due to the large coupling strength. Also  frequencies coming from transitions between the energy levels from neighboring manifolds are shown enlarged in Fig. \ref{Fig::F_e=0_D=0.5_g=2.0}. The adiabatic approximation and VVP can cover the main structure of the peaks involved there, while the former shows stronger deviations.\newline
If we go to higher values $\Delta /\Omega \gtrsim 1$, the peaks in the individual groups become more spread out in frequency space, and for the population difference dephasing already occurs at a shorter timescale. For $\Delta / \Omega =1$, at least VVP yields still acceptable results in Fourier space but gets fast out of phase for the population difference. 

\subsection{Dynamics for finite static bias $\eps \neq 0$}
\begin{figure*}
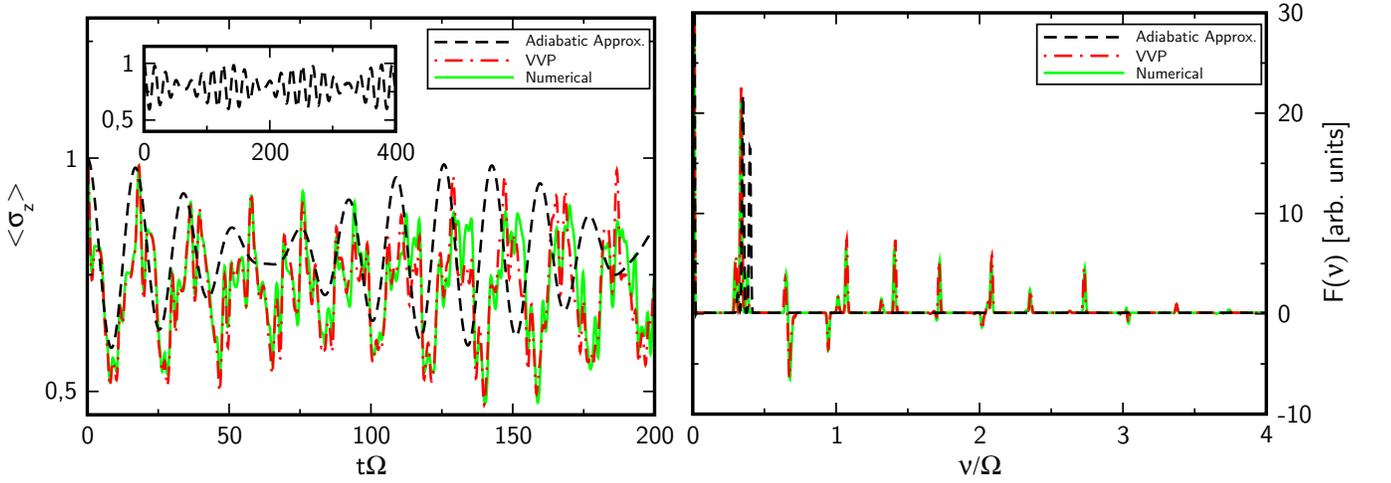

 \includegraphics[width=8.9cm]{Figures/Fig_19a_Revised.eps}
\includegraphics[width=8.9cm]{Figures/Fig_19b_Revised.eps}
 \caption{Population difference and Fourier spectrum for a biased qubit ($\eps/\Omega = \sqrt{0.5}$) at resonance with the oscillator $(\Delta_\text{b}= \Omega)$ in the ultrastrong coupling regime ($g/\Omega = 1.0$). Concerning the time evolution VVP agrees well with numerical results. Only for long time weak dephasing occurs. The inset in the left-hand figure shows the adiabatic approximation only. It exhibits death and revival of oscillations  which are not confirmed by the numerics. For the Fourier spectrum, VVP covers the various frequency peaks, which are gathered into groups like for the unbiased case. The adiabatic approximation only returns the first group.  \label{Fig::PF_e=Sqrt0.5_D=Sqrt0.5_g=1.0}}
\end{figure*}
\begin{figure*}
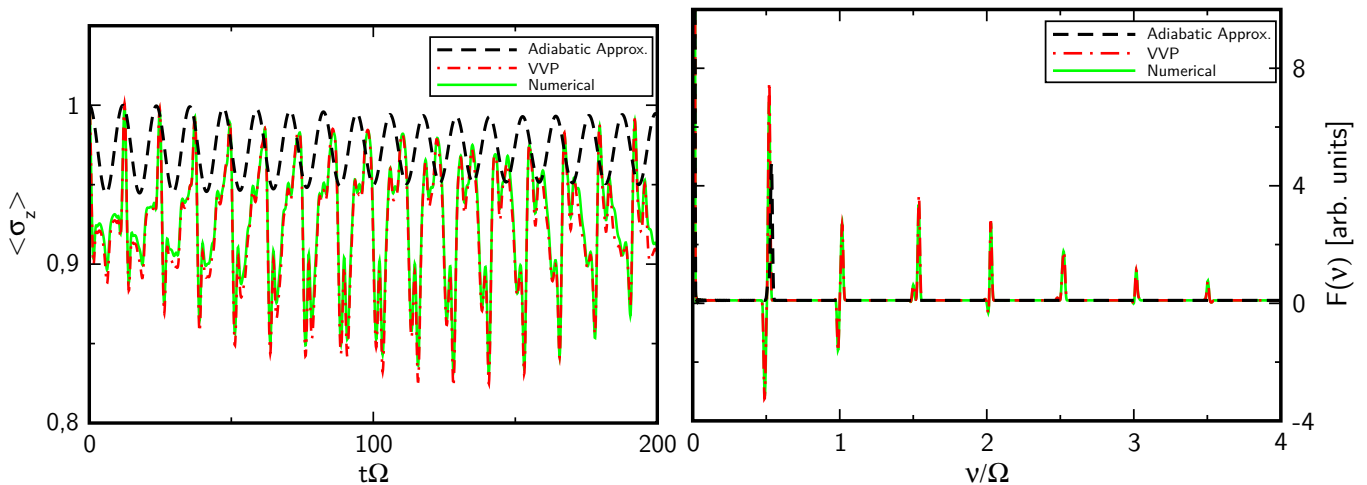

 \includegraphics[width=8.9cm]{Figures/Fig_20a_Revised.eps}
\includegraphics[width=8.9cm]{Figures/Fig_20b_Revised.eps}
 \caption{Population difference and Fourier spectrum for $\eps/\Omega = 1.5$, $\Delta/\Omega =0.5$ and $g/\Omega =1.0$. Van Vleck perturbation theory is confirmed by numerical calculations, while results obtained from the adiabatic approximation deviate strongly. In Fourier space, we find pairs of frequency peaks coming from the two dressed oscillation frequencies $\Omega_j^1$ and $\Omega_j^2$. The spacings in between those pairs is about $0.5 \Omega$. The adiabatic approximation only returns one of those dressed frequencies in the first pair.  \label{Fig::PF_e=1.5_D=0.5_g=1.0}}
\end{figure*}

As a first case, we consider in Fig. \ref{Fig::PF_e=Sqrt0.5_D=Sqrt0.5_g=1.0} a weakly biased qubit ($\eps /\Omega = \sqrt{0.5}$) being at resonance with the oscillator ($\Delta_\text{b} = \Omega$). For a coupling strength of $g/\Omega =1.0$, we find a good agreement between the numerics and VVP. The adiabatic approximation, however, conveys a slightly different picture: Looking at the time evolution it reveals collapse and rebirth of oscillations after a certain interval. This feature does not survive for the exact dynamics. Like in the unbiased case, the adiabatic approximation gives only the first group of frequencies between the quasidegenerate subspaces, and thus yields a wrong picture of the dynamics. In order to cover the higher frequency groups, we need again to go to higher-order corrections by using VVP. \newline
For the derivation of our results we assumed that $\eps$ is a multiple of the oscillator frequency $\Omega$, $\eps=l \Omega$.
In this case we found that the levels $E_{\downarrow, j}^0$ and $E_{\uparrow, j+l}^0$ form a degenerate doublet, which dominates the long-scale dynamics through the dressed oscillations frequency $\Omega_j^l$. For $l$ being not an integer those doublets cannot be identified unambiguously anymore. For instance,  we examine the case $\eps/\Omega =1.5$ in Fig. \ref{Fig::PF_e=1.5_D=0.5_g=1.0}. Here, it is not clear which  levels should be gathered into one subspace: $j$ and $j+1$ or $j$ and $j+2$. Both the dressed oscillation frequencies $\Omega_j^1$ and $\Omega_j^2$ influence the longtime dynamics.
In Fig. \ref{Fig::PF_e=1.5_D=0.5_g=1.0}, we chose $l=2$ for our approximate method. Surprisingly, VVP gives a very accurate picture for both the dynamics and the Fourier spectrum. For $l=1$ we obtained the same result (not shown here). Thus, our approach can also treat the case of $\eps$ being not a multiple of $\Omega$, and independent of the choice of $l$, VVP covers all relevant frequencies because of taking into account connections between different manifolds. We always find pairs of frequencies resulting from $\Omega_j^1$ and $\Omega_j^2$. Those pairs are separated approximately by $0.5\Omega$, which is the smallest distance between the unperturbed energy levels (only  the single levels are separated by a larger distance). For a bias of $\eps/\Omega =2.5$, for example, one would detect the same separation between the different groups of peaks. The adiabatic approximation extended to nonzero static bias fails in such a situation, as it will always only consider one of the two frequencies, which can be also seen by looking at the dynamics in Fig. \ref{Fig::PF_e=1.5_D=0.5_g=1.0}. Furthermore, as we saw already in the unbiased case, it neglects the higher frequencies for intermediate coupling.

\section{Conclusion} \label{Sec::Conclusion}
Up to now there exists no clear definition of the ultrastrong coupling regime. In many cases it is used to denote coupling strengths $g/\Omega$ for which the Jaynes-Cummings model is not valid anymore. Consequences of this failure can already be visualized for intermediate regimes like $g/\Omega \approx 0.1$ \cite{Niemczyk2010, Forn-Diaz2010, Guenter2009,Anappara2009}. At such a coupling strength it is often sufficient to take into account the counter-rotating terms in the qubit-oscillator Hamiltonian by treating it perturbatively to second order in $g$ \cite{Hausinger2008}. For stronger coupling, like  $g/\Omega$ approaching unity or going beyond, higher orders are needed. In this work we presented an approach which treats the qubit-oscillator system to all orders in $g$. The price we had to pay was to make some restriction on the tunneling matrix element $\Delta$ and thus the qubit transition frequency $\Delta_\text{b}$ compared to the oscillator frequency $\Omega$. In detail, we followed a perturbative approach with respect to the dressed tunneling element $\Delta_j^l$. However, since especially for strong coupling this dressed element becomes suppressed by a Gaussian, and by  using VVP to include also higher orders, we could go beyond the limit $\Delta \ll \Omega$ of an adiabatically fast oscillator \cite{Irish2005, Ashhab2010}. For zero bias, we compared the energy spectrum obtained by our method and the adiabatic approximation to  the generalized rotating-wave approximation in \cite{Irish2007}. For $\Delta/\Omega <1$ all approaches agree well with numerical results for the whole coupling strength, while at resonance and slight positive detuning the GRWA was found to be preferable at  weak coupling $g \to 0$, since it returns correctly the Jaynes-Cummings limit. For strong coupling and small positive detuning VVP even showed slightly better results than the GRWA. We investigated in detail the dynamics of the qubit in the zero bias case and the ultrastrong coupling regime at low temperature. While  the adiabatic approximation gives a coarse-grained picture of the time evolution of the population difference, VVP also covers the higher frequencies agreeing well with numerical results. 
For not too weak coupling our approach even gives reliable results at the resonance point $\Delta = \Omega$ and slightly beyond. \newline
The dynamics obtained in the ultrastrong coupling regime is much richer than the ones predicted by the JCM. Instead of  two dominating frequencies we found groups of peaks whose splitting is not linear in $g$ anymore as found for the common vacuum Rabi oscillations but rather depends  in a non-trivial way on a dressing by Laguerre polynomials. With the help of our analytical formulas we could understand this structure. The situation reminds of the case of a classically driven TLS, where the resulting Rabi frequency, which for weak driving is linear in the driving amplitude, shows a Bessel function like dependence in the case of extreme driving \cite{Hausinger2010}. The dressing of the qubit-oscillator system by Laguerre polynomials allows a suppression of specific frequencies through a variation of the coupling strength $g$. Finally, we could see  from the expressions (\ref{DressedOsc}) for the dressed oscillation frequency and (\ref{QubitOscillatorState}) for the second-order eigenstates that one cannot speak of single qubit or oscillator contributions anymore but has to consider a highly entangled system even for the ground state.\newline
Furthermore, we examined the situation of a biased qubit, which so far has not been treated analytically for the regime of comparable qubit and oscillator frequency ($\Delta_\text{b} \sim \Omega$). An extension of the adiabatic approximation to the biased case was almost automatically included in our treatment. We showed that for situations where the bias is not a multiple of the oscillator frequency, it is necessary to take connections between different manifolds into account. Our approach is valid at resonance as well as positive and negative qubit detuning, provided that $\Delta \lesssim \Omega$ and/or strong coupling $g/\Omega$.\newline
As we already stated above, for weak coupling strengths like $g/\Omega \sim 10^{-2}$ our approach cannot represent a replacement to the exactly solvable Jaynes-Cummings model. Also in the intermediate range  of $g/\Omega \sim 10^{-1}$, perturbative approaches or the GRWA for zero bias calculation might be preferable.
They fail, however, in the case of even stronger coupling - especially if the qubit is tuned away from its symmetry point. Here, our method shows that a new physical behavior can be expected, for which first hints have been given in recent experiments. 

\begin{acknowledgments}
 We acknowledge financial support under DFG Program SFB631.
\end{acknowledgments}

\appendix

\section{Eigenstates obtained by Van Vleck perturbation theory} \label{App::Eigenstates}
For $\Delta = 0$ the eigenstates of the qubit-oscillator system read
\begin{align} 
 \ketU{j} = & \sum_{j^\prime=0}^\infty [\sign{j-j^\prime}]^{|j^\prime - j|} \DXi{|j^\prime -j|}{\Min{j,j^\prime}}(\alpha/4) \ket{\uparrow,j^\prime}, \label{App::Delta0stateUp}  \\
 \ketD{j} = & \sum_{j^\prime=0}^\infty [\sign{j^\prime-j}]^{|j^\prime - j|} \DXi{|j^\prime -j|}{\Min{j,j^\prime}}(\alpha/4) \ket{\downarrow,j^\prime}. \label{App::Delta0stateDown}
\end{align}
The matrix elements of the Van Vleck transformation $S$ in the basis of these states  are to first order \footnote{For the general formulas and an explanation of Van Vleck perturbation theory see, e.g., \cite{Hausinger2008}.}:
\begin{align}
\bra{\widetilde{j, \downarrow / \uparrow}} \rmi S^{(1)} \ket{\widetilde{j^\prime,\uparrow / \downarrow}} = &  - \frac{(\pm)^{|j^\prime-j|}}{2} \frac{\Delta_j^{j^\prime}}{\eps \mp (j^\prime -j) \Omega} \nonumber \\
    & \times (1-\delta_{j \pm l,j^\prime}).
\end{align} 
To second order we get,
\begin{align}
 & \bra{\widetilde{\uparrow,j}} \rmi S^{(2)} \ket{\widetilde{\uparrow, j^\prime}} =  \frac{1}{4 (j^{\prime}- j)\Omega} \biggl\{ \sum_{\substack{k=0 \\ k \neq \{j - l, j^\prime - l\}}}^\infty \frac{\Delta_k^j \Delta_k^{j^\prime}}{2} \nonumber \\
                  & \times \left[ \frac{1}{ \eps+ (k-j) \Omega} + \frac{1}{ \eps+ (k-j^\prime) \Omega}\right] + \frac{\Delta_{j^\prime - l}^j \Delta_{j^\prime - l}^{j^\prime}}{  \eps + (j^\prime - l-j) \Omega} \nonumber \\
                  & + \frac{\Delta_{j - l}^j \Delta_{j- l}^{j^\prime}}{ \eps + (j- l-j^\prime) \Omega} \biggr \} (1- \delta_{j,j^\prime}),
\end{align}
\begin{align}
 & \bra{\widetilde{\downarrow,j}} \rmi S^{(2)} \ket{\widetilde{\downarrow, j^\prime}} =  \frac{1}{4 (j^{\prime}- j)\Omega} \biggl\{ \sum_{\substack{k=0 \\ k \neq \{j + l, j^\prime + l\}}}^\infty \frac{\Delta_j^k \Delta_{j^\prime}^k}{2} \nonumber \\
                  & \times \left[ \frac{1}{- \eps+ (k-j) \Omega} + \frac{1}{- \eps+ (k-j^\prime) \Omega}\right] \nonumber\\
                  &  + \frac{\Delta_j^{j^\prime + l} \Delta^{j^\prime + l}_{j^\prime}}{ - \eps + (j^\prime + l-j) \Omega} 
                   + \frac{\Delta^{j + l}_j \Delta^{j+ l}_{j^\prime}}{- \eps + (j+ l-j^\prime) \Omega} \biggr \} (1- \delta_{j,j^\prime}).
\end{align}
Using the above expressions, we find the eigenstates of $H$ to second order in $\Delta$ as
\begin{equation} \label{QubitOscillatorState}
 \ket{\Phi_{\pm, j}} = \ket{\Phi_{\pm,j}^{(0)}} + \ket{\Phi_{\pm, j}^{(1)}} + \ket{\Phi_{\pm, j}^{(2)}}
\end{equation} 
with
\begin{align}
  & \ket{\Phi_{-,j}^{(1)}} =  \sinmixa \sum_{j^\prime=0}^\infty \ketU{j^\prime} \braU{j^\prime} \rmi S^{(1)} \ketD{j} \nonumber \\
   &+ \sign{\Delta_j^{j+l}} \cosmixa \sum_{j^\prime=0}^\infty \ketD{j^\prime} \braD{j^\prime} \rmi S^{(1)} \ketU{j+l}
\end{align} 
and
 \begin{align}
  & \ket{\Phi_{-,j}^{(2)}} =  \sinmixa \sum_{j^\prime=0}^\infty \ketD{j^\prime} \braD{j^\prime} \rmi S^{(2)} \ketD{j} \nonumber \\
   &+ \sign{\Delta_j^{j+l}} \cosmixa \sum_{j^\prime=0}^\infty \ketU{j^\prime} \braU{j^\prime} \rmi S^{(2)} \ketU{j+l}  \nonumber \\
   & - \frac{1}{2} \sinmixa \sum_{j^\prime=0, k^\prime =0}^\infty \ketD{j^\prime} \braD{j^\prime} \rmi S^{(1)} \ketU{k^\prime} \braU{k^\prime} \rmi S^{(1)} \ketD{j} \nonumber \\
  & - \frac{1}{2} \sign{\Delta_j^{j+l}} \cosmixa \nonumber \\
     & \times \sum_{j^\prime=0, k^\prime =0}^\infty \ketU{j^\prime} \braU{j^\prime} \rmi S^{(1)} \ketD{k^\prime} \braD{k^\prime} \rmi S^{(1)} \ketU{j+l}.
\end{align}
For $\ket{\Phi_{+,j}^{(i)}}$ one just replaces $\sinmixa \rightarrow -\cosmixa$ and $\cosmixa \rightarrow \sinmixa$.

\section{Eigenenergies for $\eps =0$ using VVP} \label{App::VVEnE0}
We perform the summation in Eq. (\ref{VVEnE0}) and show analytical expressions for the first four energy levels obtained from VVP for the zero static bias case $\eps =0$:
\begin{align}
 E_{\mp, 0} =& \hbar \biggl[ -\frac{g^2}{\Omega} + \frac{\Delta^2\rme^{-\alpha} }{4 \Omega}  (\Gamma(0,-\alpha)+\ln(-\alpha)+\gamma)  \nonumber\\
            &\mp \frac{1}{2} |\Delta \rme^{-\alpha/2}|  \biggr],
\end{align} 
\begin{align}
 & E_{\mp,1} = \hbar \biggl[ \Omega  -\frac{g^2}{\Omega} + \frac{\Delta^2\rme^{-\alpha} }{4 \Omega} \biggl\{ 1 +\gamma + \rme^\alpha (\alpha -1) \nonumber \\
 &-\alpha [\alpha -\gamma (\alpha -2)] + (\alpha -1)^2  [\Gamma(0,-\alpha)+ \ln(-\alpha)]\biggr\} \nonumber \\
 & \mp \frac{1}{2} |\Delta (1-\alpha) \rme^{-\alpha/2}|  \biggr],
\end{align}
where we used the Euler-Mascheroni constant $\gamma$ and the incomplete $\Gamma$-function \cite{Arfken2001}. 

\section{The generalized rotating-wave approximation (GRWA)}\label{App::GRWA}
Since we use the generalized rotating-wave approximation in Sec. \ref{Sec::EnergySpecZeroBias} where we calculate the energy spectrum at $\eps=0$ as a comparison to  our VVP results, we will sketch its derivation in this appendix. A detailed description is found in \cite{Irish2007}. The first step in its derivation is to represent the qubit oscillator Hamiltonian (\ref{HTot}) in the effective basis states (\ref{FiMineff}) and (\ref{FiPluseff}), disregarding the second-order corrections in $\Delta$. Taking into account that $\Delta_j^{j^\prime} = (-1)^{|j-j^\prime|}\Delta_{j^\prime}^j$, the corresponding matrix is for the first six basis states $\{ \ket{\Phi_{\mp, j}^{(0)}} \}$ with $j=0,1,2$
\begin{equation}
	\left( \begin{array}{ccccccc}
	E_{-,0} & 0 & 0 & \hh \Delta_0^1 & -\hh \Delta_0^2 & 0 & \ldots\\
	\\
	0 & E_{+,0} & -\hh \Delta_0^1 & 0 & 0 & \hh \Delta_0^2 & \ldots\\
	\\
	0 & -\hh \Delta_0^1 & E_{-,1} & 0 & 0 & \hh \Delta_1^2 & \ldots\\
	\\
	\hh \Delta_0^1 & 0 & 0 &  E_{+,1} & -\hh \Delta_1^2 & 0 & \ldots\\
	\\
	-\hh \Delta_0^2 & 0 & 0 & -\hh \Delta_1^2 & E_{-,2} & 0 & \ldots\\
	\\
	0 & \hh \Delta_0^2 & \hh \Delta_1^2 & 0 & 0 & E_{+,2} & \ldots\\
	\vdots & \vdots & \vdots & \vdots & \vdots & \vdots & \ddots
	\end{array}
 \right).
\end{equation}
In this representation, we neglect now the remote matrix elements, which turn out to yield fast rotating contributions for $\Delta \approx \Omega$ .  A more elaborated justification is given in \cite{Irish2007}. This procedure is quite similar to the standard rotating-wave approximation and we end up again with block-diagonal matrix,
\begin{equation} 
	\left( \begin{array}{ccccccc}
	E_{-,0} & 0 & 0 & 0 & 0 & 0 & \ldots\\
	\\
	0 & E_{+,0} & -\hh \Delta_0^1 & 0 & 0 & 0 & \ldots\\
	\\
	0 & -\hh \Delta_0^1 & E_{-,1} & 0 & 0 & 0 & \ldots\\
	\\
	0 & 0 & 0 &  E_{+,1} & -\hh \Delta_1^2 & 0 & \ldots\\
	\\
	0 & 0 & 0 & -\hh \Delta_1^2 & E_{-,2} & 0 & \ldots\\
	\\
	0 & 0 & 0 & 0 & 0 & E_{+,2} & \ldots\\
	\vdots & \vdots & \vdots & \vdots & \vdots & \vdots & \ddots
	\end{array}
 \right),
\end{equation}
which is straightforwardly diagonalized. The energy of the ground state remains unchanged, namely $E_{-,0}$. The remaining levels are
\begin{align}
   E^\text{GRWA}_{\mp,j} /\hbar =& (j+\half)\Omega - \frac{g^2}{\Omega} + \frac{\Delta}{4} \rme^{-\alpha/2}(|\mathsf{L}_j^0(\alpha)|-|\mathsf{L}_{j+1}^0(\alpha)|)\nonumber\\
   & \mp \left\{\left[ \frac{\Omega}{2} - \frac{\Delta}{4} \rme^{-\alpha/2} (|\mathsf{L}_j^0(\alpha)|+|\mathsf{L}_{j+1}^0(\alpha)|)   \right]^2  \right. \nonumber\\
    & \phantom{\mp \{}\left. + \frac{\Delta^2}{4} \frac{\alpha}{j+1} \rme^{-\alpha} [\mathsf{L}_j^1(\alpha)]^2 \right\}^\half.
\end{align} 


\end{document}